\documentclass[twocolumn, aps,prb,superscriptaddress, nolongbibliography]{revtex4-2}

\usepackage{graphicx}
\usepackage{dcolumn}
\usepackage{bm}
\usepackage{siunitx}

\usepackage{color}
\usepackage{booktabs}
\usepackage{miller} 
\usepackage{wasysym} 
\usepackage{siunitx}

\usepackage{hyperref}
\definecolor{r}{rgb}{0.86,0.08,0.23}

\definecolor{blue_n}{rgb}{0.,0.3,0.5}

\hypersetup{
backref=true,
pagebackref=true,
hyperindex=true,
colorlinks=true,
breaklinks=true,
citecolor = blue_n,
urlcolor= blue_n,
linkcolor= black,
pdftitle={Optical detection of the electron spin resonances of G centers in silicon},
pdfauthor={F\'elix~Cache, Krithika V.R., Andrej Yu Kuznetsov, Tobias Herzig, Sébastien Pezzagna, Marco Abbarchi,  Isabelle~Robert-Philip, Jean-Michel~G\'erard, Guillaume~Cassabois, Vincent~Jacques and Ana\"is~Dr\'eau }
}

\newcommand{\ket}[1]{|#1\rangle}

\begin{document}

    \title{Optical detection of the electron spin resonances of G centers in silicon}

	\author{F\'elix Cache}		\affiliation{Laboratoire Charles Coulomb, Universit\'e de Montpellier and CNRS, 34095 Montpellier, France}
	\author{Krithika V.R. }		\affiliation{Laboratoire Charles Coulomb, Universit\'e de Montpellier and CNRS, 34095 Montpellier, France}
	\author{Tobias~Herzig}		\affiliation{Division of Applied Quantum Systems, Felix-Bloch Institute for Solid-State Physics, University Leipzig, Linn\'eestra\ss e 5, 04103 Leipzig, Germany}
	\author{ Andrej~Yu~Kuznetsov}  \affiliation{Department of Physics, University of Oslo, Oslo, Norway}
	\author{S\'ebastien~Pezzagna}	\affiliation{Division of Applied Quantum Systems, Felix-Bloch Institute for Solid-State Physics, University Leipzig, Linn\'eestra\ss e 5, 04103 Leipzig, Germany}
	\author{Marco~Abbarchi}	 \affiliation{CNRS, Aix-Marseille Universit\'e, Centrale Marseille, IM2NP, UMR 7334, Campus de St. J\'er\^ome, 13397 Marseille, France}
		\affiliation{Luminy Biotech, 163 avenue de Luminy, 13009 Marseille FRANCE}
	\author{Isabelle~Robert-Philip}   \affiliation{Laboratoire Charles Coulomb, Universit\'e de Montpellier and CNRS, 34095 Montpellier, France}
	\author{Jean-Michel~G\'erard}	\affiliation{Univ. Grenoble Alpes, CEA, Grenoble INP, IRIG, PHELIQS, 38000 Grenoble, France}
	\author{Guillaume Cassabois}	 \affiliation{Laboratoire Charles Coulomb, Universit\'e de Montpellier and CNRS, 34095 Montpellier, France}
		\affiliation{Institut Universitaire de France, 75231 Paris, France}
	\author{Vincent Jacques}	 \affiliation{Laboratoire Charles Coulomb, Universit\'e de Montpellier and CNRS, 34095 Montpellier, France}
	\author{Ana\"is Dr\'eau}	\email{anais.dreau@umontpellier.fr}	 \affiliation{Laboratoire Charles Coulomb, Universit\'e de Montpellier and CNRS, 34095 Montpellier, France} \email{anais.dreau@umontpellier.fr}


    \begin{abstract}
     Color centers in silicon are emerging as promising platforms for quantum technologies. Among them, the G center has attracted considerable interest owing to its bright telecom O-band single-photon emission and its optically addressable metastable electron-spin triplet state.
Here we investigate the spin properties of ensembles of G centers under above-band-gap excitation. 
We elucidate the spin photo-dynamics giving rise to the optical detected magnetic resonance (ODMR) response of G centers. 
The optimal pulsed sequence for measuring the ODMR spectrum of the G defects is identified, along with the temperature and optical-power regimes maximizing the spin readout contrast. 
Through magneto-optical measurements, we detect a level-anticrossing of the G center electron spin states.
At last, we demonstrate coherent spin control of the defects, and characterize their spin-coherence properties.
Unveiling the spin degree of freedom of the G center opens new avenues for the realization of quantum memories and quantum registers based on silicon color centers.
    \end{abstract}

    \maketitle

	\section{Introduction}

The recent isolation of single optically active defects in silicon opens new opportunities for the development of quantum technologies leveraging the mature CMOS nanofabrication processes~\cite{redjem_single_2020, durand_broad_2021,  higginbottom_optical_2022, baron_detection_2022,baron_single_2022,hollenbach_wafer-scale_2022, gritsch_purcell_2023, jhuria_programmable_2024, sandholzer_single-photon_2026}.
A key asset of these individual emitters is their telecom-band single-photon emission, which makes them particularly attractive for quantum communications and integrated quantum photonics~\cite{sandholzer_single-photon_2026, simmons_scalable_2024}.
For some defects, the electron spin degree of freedom provides also a long-lived spin qubit, whose state can be optically initialized and read~\cite{bergeron_siliconintegrated_2020, higginbottom_optical_2022, gritsch_optical_2025, cache_single_2025}. 
Additionally, electron spin single-shot readout has been demonstrated on individual integrated Er dopants~\cite{gritsch_optical_2025},
while single T centers have been harnessed to generate entanglement locally between nuclear spins~\cite{song_entanglement_2026}, and remotely between distant defects~\cite{inc_distributed_2024}.
The G center is another attractive fluorescent spin-defect in silicon, combining telecom single-photon emission at high rates \cite{kim_bright_2025, ma_nanoscale_2025, cache_single_2025} with a coherently controlled electron spin triplet~\cite{cache_single_2025}.

The potential of the G center for silicon-based quantum technologies has driven recent advances along two main complementary directions. 
On the one hand, reproducible G center engineering methods have to be developed, which include carbon and proton co-implantation \cite{baron_single_2022} and focused ion-beam techniques \cite{hollenbach_wafer-scale_2022} for the formation of single defects, and laser writing for ensembles \cite{quard_femtosecond-laser-induced_2024,andrini_activation_2024,gu_end--end_2025}.
On the other hand, understanding the optical properties of individual G-centers \cite{durand_hopping_2024,komza_indistinguishable_2024} and how to control them through integration within nano-photonic devices have made remarkable progress.
Purcell-enhanced single-photon emission of individual G centers has been reported in various photonic cavities~\cite{saggio_cavity-enhanced_2024,kim_bright_2025, ma_nanoscale_2025,cache_single_2025}.
Meanwhile, Stark effect in p-n junctions and strain engineering in micro-electromechanical cantilevers can be used to tune the G center emission energy~\cite{day_electrical_2024, buzzi_spectral_2025}.

Control of the G-center spin properties could further strengthen its potential for quantum information science and technology.
Electron spin resonance (ESR) experiments performed in the 1980s established that the G center hosts an electron-spin triplet in a metastable (MS) level, giving rise to optically detected magnetic resonance (ODMR)~\cite{lee_optical_1982,odonnell_origin_1983}.
After these pioneering studies, the spin properties of the G center remained largely unexplored until the recent demonstration of coherent control of single G-center spins \cite{cache_single_2025}.
However, these results focused on a spin-tumbling effect in single centers and did not provide a comprehensive picture of the G-center spin properties.
In particular, the photo-excited spin dynamics underlying the G center ODMR response remains poorly understood, in a material in which ODMR spectroscopy is notoriously challenging to perform~\cite{weman_optical_1988,chen_delayed_1990,chen_role_1991}.

In this manuscript, we explore the electron-spin properties of G center ensembles under above-bandgap optical excitation.
We first analyze the spin-dependent photo-dynamics of the defects through differential time-resolved photoluminescence (TRPL) experiments. 
Next, we evidence a laser-assisted deshelving mechanism impacting the spin population in the MS level of the G center.
These findings then allow us to determine the optimal pulse sequence to measure ODMR spectra on G centers with maximal spin readout contrast. 
Furthermore, we perform magneto-optical measurements and  detect a level-anticrossing of the G center electron spin states.
At last, the electron spin coherent control of the G centers is demonstrated, along with the investigation of their spin coherence times.

	\section{G center energy level structure}

			\begin{figure}[h!]
		\includegraphics[width=\columnwidth]{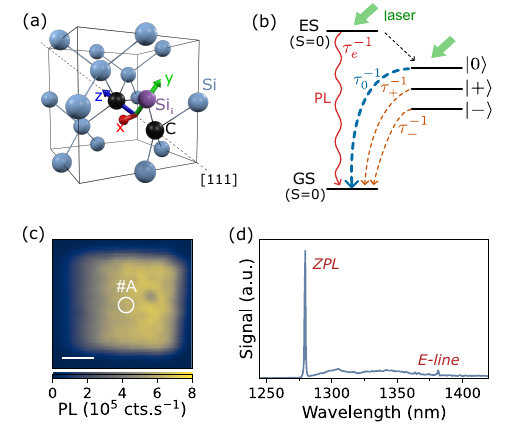}
		\caption{
		(a) Microscopic structure of the G center in silicon.
		(b) Energy level structure of the G center and relaxation rates. $\{\ket{0},\ket{+},\ket{-}\}$ are the zero-magnetic-field spin eigenstates of the metastable spin triplet level.
		(c) PL scan of the \#S1 sample.
		Scale bar is 5~\unit{\micro\meter}. 
		(d) PL spectrum of the G center ensemble \#A, recorded at \textcolor{black}{6} K.
		}
		\label{fig:intro}
	\end{figure}

\begin{figure*}
		\includegraphics[width=\textwidth]{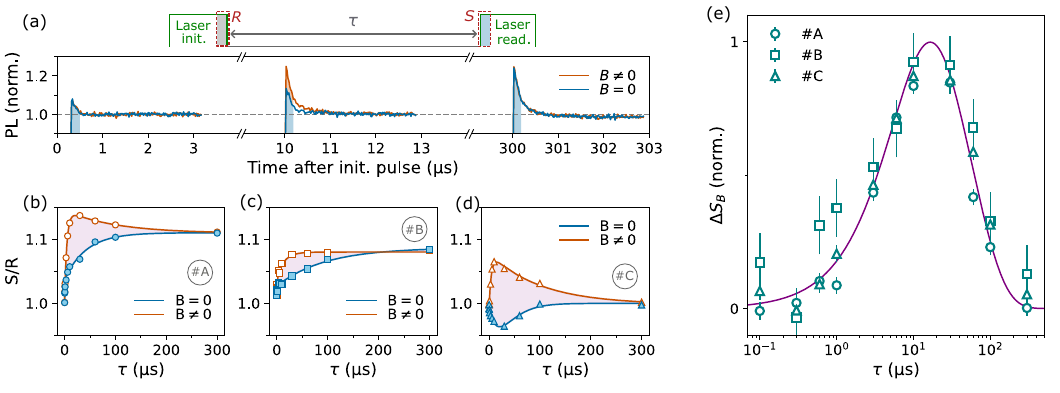}
		\caption{
		(a) TRPL traces acquired on ensemble \#A for delays $\tau$ of 0.3, 10 and 300~\unit{\micro\second}, at $B=0$ and $B = 50$~mT aligned along [001].
		Colored areas indicate the 200-ns integration window used  to infer the signal $S$ (see pulse sequence on top). 
		The normalization signal $R$ is taken at the end of the previous laser pulse.
		(b),(c),(d) For the three G center ensembles, evolution of $S/R$ versus waiting time $\tau$ in the dark.
		Solid lines are guides to the eye. 
		Optical powers: \#A 10~\unit{\micro\watt},  \#B 3~\unit{\micro\watt}, \#C 3~\unit{\micro\watt}.
		(e) Normalized difference signal $\Delta S_B = S/R_{B \neq 0} - S/R_{B=0}$ for the three G ensembles. 
		Solid line is the result of data fitting with $a\left(e^{-\tau/\tau_1} - e^{-\tau/\tau_\text{eff}}\right)$ (see text).
		}
		\label{fig:MS}
	\end{figure*}

		The G center in silicon is composed of two substitutional carbon atoms bound to one interstitial silicon atom Si$_{\mathrm{i}}$ (Fig.~\ref{fig:intro}(a)).
	This defect can exhibit an uncommon reorientation dynamics driven by the motion of this Si$_{\mathrm{i}}$ atom between 6 sites~\cite{odonnell_origin_1983,udvarhelyi_identification_2021,durand_hopping_2024}.
Its energy level structure is comparable to that of fluorescent molecules with an optical transition between ground (GS) and excited (ES) states with $S=0$, and an intermediate MS electron spin triplet level $S=1$ (Fig.~\ref{fig:intro}(b)).
	Under above-bandgap optical excitation in unstructured silicon samples, the G center can relax radiatively from the ES on a time-scale $\tau_e \simeq 5$~\unit{\nano\second}~\cite{beaufils_optical_2018}.
	The MS can be populated either from the ES by inter-system crossing transitions, or by direct transitions from the conduction and valence states.
	At zero magnetic field, the degeneracy between the MS electron spin triplet sublevels is lifted by spin-spin interaction, described by the zero-field splitting (ZFS) Hamiltonian: 
	\begin{equation}
	\hat{\mathcal{H}}_0=D\hat{S}_z^2 + E(\hat{S}_x^2-\hat{S}_y^2),
	\label{eq:H}
	\end{equation}
where $D$ and $E$ denote the longitudinal and transverse ZFS components, and $\mathbf{\hat{S}}$ is the spin operator whose orientation is defined with respect to the defect symmetry.
The  spin axis  $\mathbf{z}$ is aligned with the two carbon atoms along the $\langle111\rangle$ axis, while $\mathbf{x}$ is perpendicular to the plane formed by the three atoms of the defect~\cite{lee_optical_1982}. 
The spin eigenstates of $\hat{\mathcal{H}}_0$ are: \{${{|0\rangle}, |+\rangle, |-\rangle}$\}, where ${|0\rangle = |m_s = 0 \rangle}$ and ${|\pm\rangle = (|m_s\!=\!+1\rangle\pm|m_s\!=\!-1\rangle)/\sqrt{2}}$, with \{$|m_s \rangle$\} the eigenstates of $\hat{S}_z$.
Their respective energies are $\{{0, D+E, D-E}\}$, with $D<0$ and $E>0$~\cite{udvarhelyi_identification_2021}.
	According to Ref. \cite{udvarhelyi_identification_2021}, the only transition allowed at first-order by spin-orbit coupling between the MS  and the GS  is the one from the $|0\rangle$ spin state.
	Consequently, the population in this spin state relaxes much faster than the one in the two other spin states $|\pm\rangle$: i.e. $\tau_0 \ll \tau_+, \tau_-$ (Fig.~\ref{fig:MS}(a)).
	This spin-dependent population relaxation underlies the ODMR response of the electron spin triplet of the G center.

\section{Samples and methods}

To investigate the G center spin-dependent photodynamics while benefiting from a high photoluminescence (PL) signal, we focus on dense ensembles of G defects created by co-implantation with carbon ions and protons \cite{berhanuddin_co-implantation_2012}.
Two silicon-on-insulator (SOI) samples are used: \#S1 coming from a commercial 220-nm SOI wafer (Soitec) \cite{beaufils_optical_2018}, and \#S2 having a $\simeq 60$-nm thick top layer of isotopically purified $^{28}$Si~\cite{baron_single_2022}. 
\#S1 and \#S2 have been implanted with carbon fluences of $5 \times 10^{13}$ and $3 \times 10^{12}$ cm$^{-2}$, respectively. 
After a 20-s rapid thermal annealing at 1000$^{\circ}$C in N$_2$ atmosphere, both samples have been locally irradiated with proton fluences of $3 \times 10^{14}$ and $3\times 10^{13}$~cm$^{-2}$, respectively. 
Three G center ensembles are investigated here: \#A in the first sample, and \#B and \#C in the second sample at $\simeq 5$~\unit{\micro\meter} apart.

For this study, the samples are placed in a home-made low-temperature confocal microscope built up in a He closed-cycle cryostat equipped with 3D superconducting Helmholtz-coils.
	Unless otherwise specified, the temperature is maintained between 4 and \SI{8}{\kelvin}.
	The optical excitation is performed using \SI{10}{\micro\watt} of a 532~nm laser focused onto the sample using a microscope objective with a numerical aperture of 0.80.
	The sample PL is collected by the same objective and detected by a superconducting nanowire single-photon detector with an efficiency of 78\% at \SI{1.3}{\micro\meter}.
	The PL signal is filtered in the 1250-1450~nm range. 
	The microwave (MW) magnetic field used to drive the electron spin of the G center is delivered to the sample through a copper wire spanning it surface.

Optical scans performed on sample \#S1 show a high PL signal in the area that has been locally irradiated with protons (Fig.~\ref{fig:intro}(c), see SI for \#S2). 
As shown in Figure~\ref{fig:intro}(d), the PL spectrum recorded on ensemble \#A exhibits the G center emission fingerprint: a zero-phonon line (ZPL) at 1279 nm, followed by a broad phonon sideband that includes the E-line at 1380~nm, associated with a localized phonon mode~\cite{davies_optical_1989, beaufils_optical_2018, durand_genuine_2024}.

\section{Spin-dependent metastable level lifetimes}\label{sec:MS}

	\begin{figure}[h!]
		\includegraphics[width=\columnwidth]{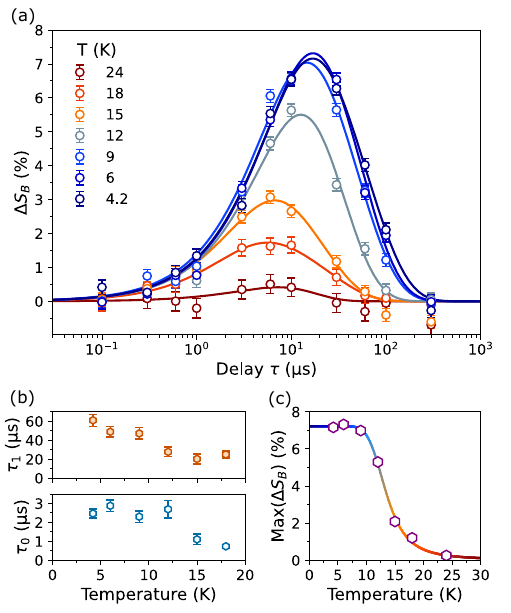}
		\caption{
		(a) Experimental results of the differential TRPL pulse sequence with increasing temperature.
		Solid curves are data fitting results with $a(e^{-\tau/\tau_1}-e^{-\tau/\tau_{eff}})$.
		(b) Evolution of $\tau_1$ and $\tau_0$ with temperature (see main text).
		(c) Maximum value of data fitting from (a) versus temperature.
		Solid line represents the fitting results with an Arrhenius law providing activation energy: $E_a = 8.7(7)$~meV.}
		\label{fig:MS_T}
	\end{figure}

The design of optimized pulse sequences for spin detection and control requires knowledge of the G-center spin-dependent MS lifetimes.
Such lifetimes are typically measured from the recovery of the PL signal after a variable waiting time in the dark \cite{robledo_spin_2011}.
The pulse sequence used for this measurement consists of two identical laser pulses for initialization and readout, separated by a variable delay $\tau$.
The first laser pulse populates the MS.
During $\tau$, its three sublevels decay to the GS with spin-dependent rates.
The PL signal at the beginning of the readout laser pulse probes the population that has relaxed to the GS during the dark time.
PL time traces recorded on ensemble \#A for different waiting times are shown in Figure~\ref{fig:MS}(a). 
A sharp PL peak appears at the beginning of the readout laser pulse, whose intensity increases with waiting time, consistently with the progressive relaxation of the MS population to the GS.
In Figure~\ref{fig:MS}(b-d), we show, for ensembles \#A, \#B, and \#C, the evolution of the PL signal $S/R$, where $S$ and $R$ are integrated in a 200-ns window at the beginning and at the end of the laser pulses. 
The dynamics vary strongly from one ensemble to the other, even within the same sample.
This discrepancy cannot be explained solely by population relaxation from the G-center MS and shows the contribution of parasitic PL signals.
In particular, the PL decrease observed in \#C is likely associated with conformational and/or charge-state fluctuations occurring in the dark \cite{song_bistable_1988}.

In order to probe the spin-dependent photo-dynamics, we repeat the experiment under a 50~mT magnetic field~$\mathbf{B}$, applied along the [001] silicon crystal axis (Fig.~\ref{fig:MS}(a-d)).
The transverse magnetic component mixes the spin states through the Zeeman Hamiltonian $\hat{\mathcal{H}}_{\mathbf{B}} = -\gamma_e \mathbf{B}\cdot \hat{\mathbf{S}}$, with $\gamma_e \simeq -28.0$~\unit{\mega\hertz/\milli\tesla}.
At short and long delays, the PL signals measured with and without magnetic field are identical.
At intermediate $\tau$, the luminescence intensity emitted by the three G ensembles is stronger for $\mathbf{B}\neq 0$, in line with an overall shortening of the MS relaxation times induced by the mixing of the three spin states. 
Nevertheless, here again, each of the three ensembles exhibits a radically different PL signal temporal evolution (Fig.~\ref{fig:MS}(b-d)).

To isolate the intrinsic photo-dynamics associated with the G center MS spin triplet, we eliminate the extrinsic parasitic contribution by computing the difference between PL intensities recorded with and without magnetic field. 
Figure~\ref{fig:MS}(e) shows the resulting differential signal $\Delta S_B= S/R_{B \neq 0} - S/R_{B=0}$.
The data associated with the three ensembles are now superimposed on the same bell-shaped curve, revealing an underlying spin photo-dynamics intrinsic to the G centers.
To extract the MS spin-dependent lifetimes, we consider that, at $B=0$, the dynamics is dominated by the relaxation of the two long-lived MS spin states $|\pm\rangle$, associated with a mean lifetime $\tau_1$, with $\tau_1^{-1} = ( \tau_{+}^{-1} +  \tau_{-}^{-1})/2 $.
Assuming complete mixing of the three spin states under perpendicular magnetic field, the effective MS decay rate reads:
\begin{equation}
\frac{1}{\tau_{\text{eff}}} \simeq \frac{1}{3}\left( \frac{1}{\tau_0} + \frac{2}{\tau_1}\right),
\label{eq:tau_eff}
\end{equation}
where $\tau_0$ denotes the lifetime associated with the MS spin state~$\ket{0}$ (see SI for rigorous calculation).
The lifetimes are thus inferred from data fitting with the bi-exponential function $a\left(e^{-\tau/\tau_1} - e^{-\tau/\tau_\text{eff}}\right)$, yielding $\tau_\text{eff}=7(2)$~\unit{\micro\second} and $\tau_1=55(10)$~\unit{\micro\second}. 
Substituting these parameters into Eq.~(\ref{eq:tau_eff}) gives $\tau_0=2.5(6)$~\unit{\micro\second}, so about $20 \times$ shorter than $\tau_1$. 
The strong difference between the $|\pm\rangle$ and $|0\rangle$ MS lifetimes is key to enable optical spin polarization and read-out of G centers in silicon.

\label{sec:MS_T}

We next investigate the temperature dependence of the G center MS lifetimes, using the same differential protocol. 
As the temperature increases from 4  to 24~K, the amplitude of $\Delta S_B$ decreases, while its maximum is shifted towards shorter delays (Fig.~\ref{fig:MS_T}(a)). %
Both $\tau_0$ and $\tau_1$ lifetimes decrease by a factor of roughly 3 as the temperature rises (Fig.~\ref{fig:MS_T}(b)).
On the other hand, the maximum of the $\Delta S_B$ signal exhibits a stronger temperature dependence: it remains constant up to 9~K and then declines rapidly, following an Arrhenius law with activation energy $E_a =8.7(7)$~meV (Fig.~\ref{fig:MS_T}(b)).
This activation energy is consistent with earlier EPR measurements from the 1980s, which attributed the signal drop to a change of the G defect's symmetry  induced by a thermally-activated reorientation \cite{odonnell_origin_1983, afanasjev_thermally_1997}.

\section{Photo-detrapping from the metastable spin triplet level}
\label{sec:detrap}

				\begin{figure}
		\includegraphics[width=\columnwidth]{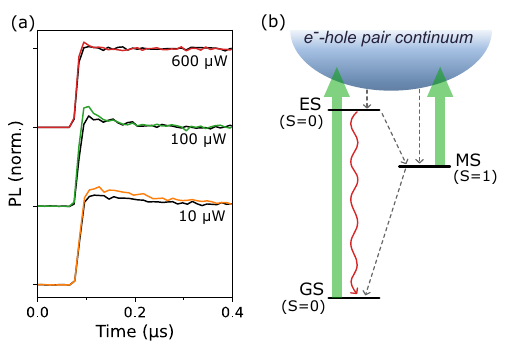}
		\caption{
		(a) TRPL signal with a 1-\unit{\micro\second} laser pulse every 15 \unit{\micro\second}, measured at different optical excitation powers for $B=0$ (black) and $B=50$~mT along $[001]$ crystal axis (color).
		(b) Simplified energy level structure showing a possible photo-detrapping mechanism induced by photoionization under green laser excitation.}
		\label{fig:detrap}
	\end{figure}

Because the spin-dependent lifetimes $\tau_0$ and $(\tau_+, \tau_-)$ differ by more than one order of magnitude, one might expect
a strong variation of the G-center PL signal under a high magnetic field applied perpendicular to the defect main spin axis.
However, $\Delta S_B$ does not exceed a few percent in practice (Fig.~\ref{fig:MS}(e)).
To elucidate the origin of this relatively weak magneto-optical response, we investigate the mechanisms responsible for populating the G center MS level.
This is achieved by analyzing the evolution of the TRPL traces with increasing optical excitation power, for a 15-\unit{\micro\second} interpulse delay chosen to maximize $\Delta S_B$ (Fig.~\ref{fig:MS}(e)).
TRPL  traces recorded on the G center ensemble, while filtering here in the 1275-1300 nm range to minimize parasitic PL contribution, are shown in Figure~\ref{fig:detrap}(a).
With increasing optical power, the PL overshoot at the start of the laser pulse gets narrower, due to the acceleration of the photo-dynamics at higher pumping rate~\cite{robledo_spin_2011}. 
But most importantly, the overshoot amplitude decreases and vanishes at high power, typically $>$~500  \unit{\micro\watt} (Fig.~\ref{fig:detrap}(a)). 
Such effect highlights the presence of an optically-induced deshelving rate impacting the MS population. 
When this rate becomes faster than the MS pumping rate, the MS is not populated anymore.
A similar effect was also evidenced in measurements of the second-order correlation function of single G* and W centers in silicon through a strong reduction of photon bunching when laser power increases~\cite{fleury_nonclassical_2000,redjem_single_2020}.
A possible deshelving mechanism, illustrated in Figure~\ref{fig:detrap}(b), could involve a photo-induced transition from the MS to the electron-hole pair continuum associated with the conduction and valence bands.

The photo-detrapping mechanism under above-band gap excitation has a strong impact on the spin photo-dynamics of the G center in silicon.
First, ODMR requires operation at low optical excitation power, thus limiting the G defect PL signal. 
Hence, for single-spin studies, the integration of individual G centers in photonic cavities to enhance their emission rate is compulsory~\cite{cache_single_2025}. 
Secondly, even at this low optical power, the photo-detrapping rate is much higher than the relaxation rates from MS to GS, including the strongest decay rate from $|0\rangle$. 
Consequently, spin initialization is not possible under continuous green above-bandgap optical pumping for the G center in silicon.
Instead, as we will see in the next section, spin state polarization can be achieved under pulsed optical excitation by relying on appropriate waiting times.

\section{ODMR spectra}

Knowing the MS spin triplet lifetimes enables the design a TRPL pulse sequence suited to measuring the ODMR spectrum of G centers. 
The ODMR sequence is composed of two 1-\unit{\micro\second} laser pulses, a 120-ns MW pulse in between and with an inter-pulse delay of 5~\unit{\micro\second} (Fig.~\ref{fig:odmr}(a)).
The first laser pulse is used to populate the MS spin levels equally~\cite{vlasenko_electron_1995}.
As $\tau_0 < 5$~\unit{\micro\second}$\ll\tau_1$, the waiting time enables spin polarization by conserving population mostly into the long-lived $|\pm\rangle$ states.
The ODMR spectrum recorded on G ensemble \#A is presented in Figure~\ref{fig:odmr}(a).
When the frequency of the MW magnetic field pulse is resonant with one of the two $|\Delta m_s| = 1$ spin transitions, populations are transferred from $|\pm\rangle$ to the short lived state $|0\rangle$, resulting in an increase of the defect emission counts by roughly 1\% (Fig.~\ref{fig:odmr}(a)).

We observe two ESR lines at $\nu_+ = $~\SI{690 \pm 10}{\mega\hertz} and $\nu_- = $~\SI{1730\pm10}{\mega\hertz}, corresponding to the spin transitions $|+\rangle\leftrightarrow|0\rangle$ and $|-\rangle\leftrightarrow|0\rangle$, respectively. 
The linewidth discrepancy results from frequency-dependent MW attenuation inducing stronger MW power broadening on the low frequency spin transition~\cite{dreau_avoiding_2011}.  
We note that the $|+\rangle \leftrightarrow |-\rangle$ spin transition is also allowed but cannot be detected with the current pulse sequence as $\tau_+\simeq \tau_-$, resulting in negligible ODMR contrast. 
Using the signs of the ZFS parameters from~\cite{udvarhelyi_identification_2021}, we deduce $D = -(\nu_- +\nu_+)/2 = $\SI{-1210 \pm 10}{\unit{\mega\hertz}} and $E = (\nu_- -\nu_+)/2 = $\SI{520 \pm10}{\unit{\mega\hertz}}.
These ZFS values are in agreement with recent ODMR results on single G centers \cite{cache_single_2025} and former ESR ensemble measurements from the 1980s \cite{lee_optical_1982}.
Since ${D_x=-D/3+E =}$~\SI{920\pm10}{\unit{\mega\hertz}} and {$D_z = 2D/3 = $\SI{-810\pm10}{\unit{\mega\hertz}}, we note that the G center in silicon does not follow the usual ESR convention stating that the principal spin axis $z$ is oriented along the direction of the largest ZFS tensor component \cite{weil_electron_2006}.

	\begin{figure}
		\includegraphics[width=\columnwidth]{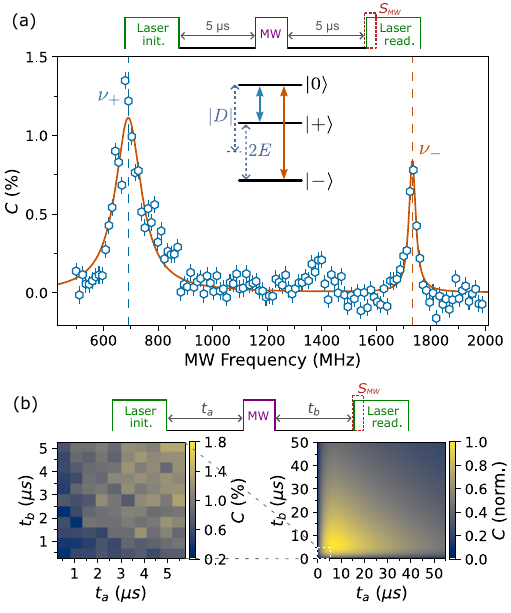}
		\caption{
		(a) ODMR spectrum measured on the G center ensemble at B~=~0~mT, using the pulsed sequence displayed on top with \SI{1}{\micro\second} laser pulses, \SI{5}{\micro\second} waiting time and \SI{120}{\nano\second} MW pulse.
		The ODMR contrast $C$ is calculated by dividing the $S_{MW}$ signal integrated on the first 200 ns of the laser pulse, by the out-of-resonance value.
		Solid line is the result of data fitting  with the sum of two Lorentzian functions.
		(b) Evolution of $C$ of the waiting times preceding ($t_a$) and following ($t_b$) the MW pulse. MW frequency was set to the   $\ket{+}\leftrightarrow\ket{0}$ transition. \textit{Left}: experimental data. \textit{Right}: simulation performed using the MS lifetimes provided in Sec.~\ref{sec:MS}.
		}
		\label{fig:odmr}
	\end{figure}

After identifying the ESR transitions, we investigate the influence of the interpulse delays on the spin readout contrast. 
Figure~\ref{fig:odmr}(b) displays the evolution of ODMR contrast $C$ for the low-frequency ESR line as a function of the delays preceding ($t_a$) and following ($t_b$) the MW pulse. 
We observe experimentally that the ODMR contrast increases with both delays, being close to 0.2\% for $t_a=t_b=0.5$~\unit{\micro\second} and exceeding  1\% for $t_a=t_b=5$~\unit{\micro\second}.
The dependence on $t_a$ is  a consequence of the photo-detrapping effect highlighted in Sec.~\ref{sec:detrap}:  after the initialization laser pulse, the spin population is unpolarized, so preparation into the long-lived spin states must rely on the waiting time.
Following the MW pulse that transfers population into the short-lived state $\ket{0}$, a sufficiently long delay $t_b$ is required to enable relaxation to the ground state before readout.
Our experimental data yield a maximum contrast for $t_a=t_b\simeq 5$~\unit{\micro\second} (Fig.~\ref{fig:odmr}(b) \textit{left}).
This observation agrees with simulations performed using the MS lifetimes estimated in Sec.~\ref{sec:MS} (Fig.~\ref{fig:odmr}(b) \textit{right}, see SI for details).
For longer waiting times, the ODMR contrast decreases because of reduced population in the MS $|\pm\rangle$ states.

\section{Magneto-optical spin spectroscopy}

	The orientational degeneracy between spin transitions can be lifted by Zeeman effect. 
	For each of the four possible $\langle 111 \rangle$-directions of $\mathbf{z}$, there are six possible sites for the Si$_{\mathrm{i}}$ atom.
	For perfectly symmetric G centers, this corresponds to three inequivalent $\mathbf{x}$ orientations $\langle \bar{1}10\rangle$, hence $12$ possible spin orientations (Fig.~\ref{fig:intro}(a))~\cite{lee_optical_1982,odonnell_origin_1983}.
	At $B$=0, as the spin transitions are at the same energy for all G center orientations, a single resonance is observed for $|+\rangle \leftrightarrow |0\rangle$ (Fig.~\ref{fig:odmr_B}(a)).
	When applying a magnetic field $\mathbf{B}$ along $[111]$, this ODMR line splits into three, each line corresponding to a family of equivalent spin orientations with respect to $\mathbf{B}$ (Fig.~\ref{fig:odmr_B}(b)).
	The ESR line shifting towards lower frequencies corresponds to the three configurations of [111]-oriented G centers, whose $\mathbf{z}$ axis is parallel to $\mathbf{B}$.
	The two lines shifting towards higher frequencies are associated with the G center $\mathbf{z}$-orientations $[\bar{1}11], [11\bar{1}]$ or $[\bar{1}1\bar{1}]$, experiencing a strong perpendicular magnetic field tilted by the spherical angles ($\theta\simeq71^\circ$, $\phi=-90^\circ$) (degeneracy: ~3) and ($\theta\simeq71^\circ$, $\phi=30^\circ$) (degeneracy:~6)  in their spin axis coordinate system (Fig.~\ref{fig:odmr_B}(a)).
	The spin transition frequencies derived from the total Hamiltonian $\hat{\mathcal{H}}_0+ \hat{\mathcal{H}}_B$  for these three families perfectly match the magnetic-field dependence of the experimental ESR frequencies (red dashed lines in Fig.\ref{fig:odmr_B}(b)).
	The evolution of the spin eigenstates and energies with magnetic field can be found in SI for the different G center orientations.

\begin{figure}
		\includegraphics[width=\columnwidth]{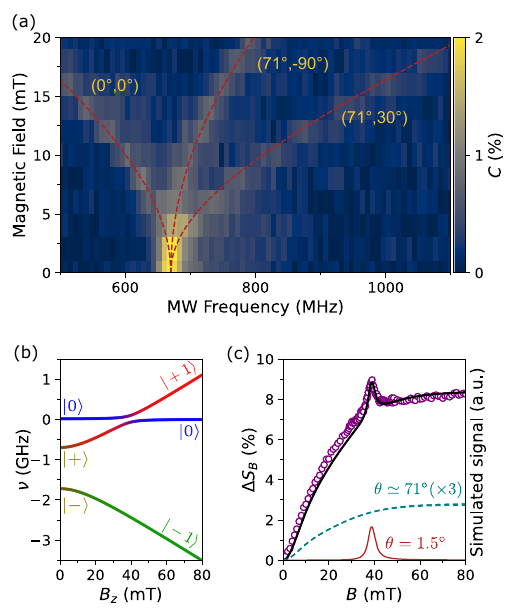}
		\caption{
		(a) Evolution of the lower branch spin resonance lines with a magnetic field $\mathbf{B}$ aligned along [111].
		The dotted red lines represent the spin transition frequencies calculated from diagonalizing $\hat{\mathcal{H}}_0+ \hat{\mathcal{H}}_B$ for the three G center families of inequivalent crystal orientations.
		(b) Evolution of the spin eigenenergies with a longitudinal magnetic field $B_z$ and an arbitrary perpendicular perturbation.
 		The blue, red and green colors are related to the spin population in states $|0\rangle$, $|\!+\!1\rangle$ and $|\!-\!1\rangle$, respectively.
 		(c) Impact of a magnetic field $\mathbf{B}$ along $[111]$ on the  TRPL signal for a 10-\unit{\micro\second} interpulse delay (markers).
		The curves show a simulation of the PL signal evolution, considering a mis-alignement angle of 1.5$^{\circ}$ between $\mathbf{B}$ and $[111]$, for the $[111]$-oriented defects (brown solid), the families at $\theta \simeq 71^{\circ}$ from $\mathbf{B}$ (teal dashes) and all orientations together (black solid). 
		}
		\label{fig:odmr_B}
	\end{figure}

	While increasing the parallel magnetic field amplitude, the spin states of the $[111]$-oriented G centers become the eigenstates of  $\hat{S}_z$:  ${|\text{m}_\text{s}\rangle}= {\{ |0\rangle, |\!+\!1\rangle, |\!-\!1\rangle\}}$ (Fig.~\ref{fig:odmr_B}(b)).
	At a magnetic field $B_{LAC} = \sqrt{D^2-E^2}/|\gamma_e| \simeq$ \SI{39 \pm 1}{\unit{\milli\tesla}}, the states $|0\rangle$ and $|\!+\!1\rangle$ have the same energy. 
	Consequently, any small perpendicular perturbation, such as hyperfine interaction or non-perfectly parallel magnetic field, leads to a level anticrossing (LAC) mixing $|0\rangle$ and $|\!+\!1\rangle$ states (Fig.~\ref{fig:odmr_B}(b)) \cite{epstein_anisotropic_2005}.
	To investigate this MS-LAC, we record the G ensemble counts while sweeping the $B$ amplitude along $[111]$, using the TRPL sequence depicted in Figure~\ref{fig:MS} with a fixed delay of 10~\unit{\micro\second}.
	The PL signal slowly increases with $B$, reaching $\simeq7\%$ at 35 mT.
	It then rises quickly to a maximum of roughly $9\%$ at  \SI{38.8\pm 0.4}{\unit{\milli\tesla}}, and finally decreases to a plateau at $\simeq 8\%$.
	The gradual rise of the signal towards this value originates from G-center families not aligned with the field, for which the spin states are rapidly mixed by the large perpendicular field component, $B \sin(\theta)$ with $\theta \simeq 71^{\circ}$.
 	Indeed, away from $B_{LAC}$, the emission rate of the G center family with $\mathbf{z} \parallel [111]$ does not change as they experience a quasi-parallel magnetic field (Fig.~\ref{fig:odmr_B}(c)).
	However, close to the MS-LAC, their spin states $|0\rangle$ and $|\!+\!1\rangle$ can be mixed by any small perpendicular perturbation, resulting in the sharp PL rise at \SI{38.8\pm 0.4}{\unit{\milli\tesla}}, in good agreement with $B_{LAC}$.
	This behavior is well reproduced by a simulation of the defects' PL signal evolution while considering a mis-alignment angle $\theta = 1.5^{\circ}$ between $\mathbf{B}$ and $[111]$  (Fig.~\ref{fig:odmr_B}(c), see SI for details).
 	Such a MS-LAC is particularly relevant to polarize and readout the states of nuclear spins coupled by hyperfine interaction to the electron spin of the defects \cite{jacques_dynamic_2009,lee_readout_2013}.

\section{Coherent control and coherence times}

 	\begin{figure}
		\includegraphics{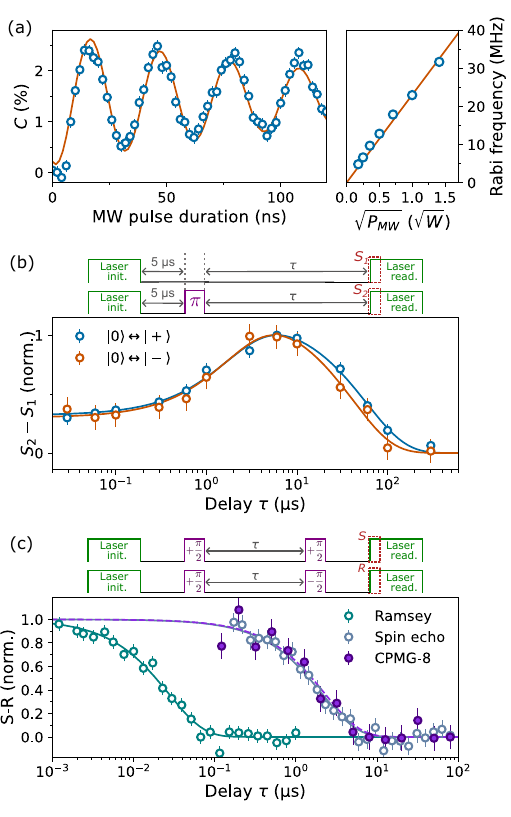}
		\caption{
		(a) Rabi oscillations measured with the sequence displayed on top.
		The solid red line represents data fitting with a sum of sine functions modulated by an exponential $e^{-t/\tau_R}$, giving $\tau_R=$~\SI{230\pm20}{\nano\second}.
		\textit{Right panel:} Evolution of the Rabi frequency  with respect to the square root of the MW power $P_{\text{MW}}$. 
		(b) Results of the differential time-resolved pulse sequence used to extract the lifetimes of the three MS spin states using the fitting function $|a_{\pm}|e^{-\tau/\tau_\pm}- |a_0| e^{-\tau/\tau_0}$:  $\tau_0=$~\SI{1.9\pm0.1}{\micro\second}, $\tau_+=$~\SI{54\pm4}{\micro\second} and $\tau_-=$~\SI{42\pm5}{\micro\second}.
		(c) Results of Ramsey, spin echo and CPMG-8 sequences. 		
		Data fitting with mono-exponential functions gives $T_2^*=$~\SI{27\pm2}{\nano\second},  $T_2^\text{echo}=$~\SI{2.1\pm0.2}{\micro\second} and $T_2^\text{CPMG-8}=$~\SI{2.3\pm0.3}{\micro\second}. 
	The Ramsey sequence is displayed on top (MW $\pi/2$ pulse: 7.2 ns).
		}
		\label{fig:rabi}
	\end{figure}

	To demonstrate coherent control of the G center electron spin we perform Rabi oscillation experiments.
	To do this, the same pulse sequence as for ODMR spectra is used but with MW pulse duration varying from 0 to 300 ns (Fig.~\ref{fig:rabi}(a)).
	The MW frequency was set resonant to the $|+\rangle\leftrightarrow|0\rangle$ spin transition.
	As displayed on Figure~\ref{fig:rabi}a, the G center fluorescence signal of ensemble \#A oscillates between low and high PL intensity, with an amplitude that decreases exponentially on a characteristic time $\tau_R=$~\SI{230\pm20}{\nano\second}.
	The Rabi frequency increases linearly with the square root of the MW power, confirming coherent driving of the G-center electron spins (Fig.~\ref{fig:rabi}(a)).

	From the Rabi oscillations, we can extract the MW $\pi$ pulse duration to get more accurate estimates of the MS spin-dependent lifetimes.
	Similar to the differential TRPL method used in Section~\ref{sec:MS}, the first sequence is made up of two laser pulses spaced 5\unit{\micro\second}$\,+\,\tau$ apart (Fig.~\ref{fig:rabi}(b)). 
	The corresponding PL signal $S_1$ measures the relaxation from the two long-lived states $|\pm\rangle$.
	For the second pulse sequence, a MW $\pi$ pulse is inserted after the 5-\unit{\micro\second} waiting time (Fig.~\ref{fig:rabi}(b)). 
	If resonant with the lower ESR frequency, the signal $S_2$ will probe the population decay from $\ket{0}$  and $\ket{-}$. 
	Computing $S_2 - S_1$ allows the parasitic signal to be eliminated and the lifetimes $\tau_0$ and $\tau_+$ to be recovered. 
	Conversely, the lifetimes $\tau_0$ and $\tau_-$ will be extracted if the MW pulse addresses the other spin transition $\ket{-}\leftrightarrow\ket{0}$. 
	The results of this differential sequence applied to the two spin transitions are presented in Figure~\ref{fig:rabi}(b).
Fitting the data with the bi-exponential function $(|a_\pm|e^{-\tau/\tau_\pm}-|a_0|e^{-\tau/\tau_0})$ yields $\tau_+ = 42(5)$~\unit{\micro\second} and $\tau_- = 54(4)$~\unit{\micro\second}, and $\tau_0 = 1.9(1)$~\unit{\micro\second} for both spin resonances.
These lifetime values are in good agreement with the estimates obtained in Section~\ref{sec:MS}.
The observation of two long and one short MS lifetimes  reinforces the assignment of $\ket{\pm}$ and $\ket{0}$ as long-lived and short-lived states, respectively.

	The coherence properties of the electron spin transitions are then investigated.
	We first measure the inhomogeneous spin dephasing time $T_2^\star$ by using  the Ramsey sequence depicted in Figure~\ref{fig:rabi}(c).
	Compared to the Rabi sequence, the MW pulse is replaced by two MW $\pi/2$ pulses separated by a varying delay.
	Here, the reference sequence used for PL signal normalization is the same, except for a $\pi$-phase shift applied on the second MW $\pi/2$ pulse.
	Data-fitting with a mono-exponential function yields $T_2^\star=$~\SI{27\pm2}{\nano\second}, thus roughly 30 times shorter than for single G centers~\cite{cache_single_2025}.
	Switching to an isotopically purified $^{28}$SiOI sample to eliminate the $^{29}$Si nuclear spin bath does not improve $T_2^\star$ (see SI).
	This suggests that electron spin decoherence of these G center ensembles is dominated by another source of noise, most likely an electron-spin bath~\cite{barry_sensitivity_2020}.
	
	To eliminate the spin dephasing related to slow magnetic fluctuations of the environment, dynamical decoupling sequences are applied to the G defect spins. 
	First, we implement the Hahn echo sequence, where a $\pi_y$ pulse is inserted at midpoint between the two $\pi_x/2$ pulses (Fig. \ref{fig:rabi}(c)).
	This refocusing pulse enables an increase of the coherence time by almost 2 orders of magnitude, up to $T_2^\text{echo}=$~\SI{2.1\pm0.2}{\micro\second}.
This coherence time cannot be further extended by adding more $\pi$ pulses, as the dynamical decoupling CPMG-8 sequence \cite{meiboom_modified_1958} leads to $T_2^\text{CPMG-8}=$~\SI{2.3\pm0.3} {\micro\second} (Fig. \ref{fig:rabi}(c)).
	This implies that magnetic noise is not the dominant source of electron spin relaxation on the $|\pm\rangle \leftrightarrow |0\rangle$ spin transitions, but rather the lifetime $\tau_0 = 1.9(1)$~\unit{\micro\second} of the short-lived MS state $|0\rangle$.

\section{Conclusion and outlook}

In conclusion, the MS electron spin photo-dynamics of G center ensembles in silicon are investigated through TRPL  magneto-optical measurements.
The MS lifetimes of the three spin states $|0\rangle$, $|+\rangle$ and $|-\rangle$ of the G center are $\tau_0 = 1.9(1)$~\unit{\micro\second},   $\tau_+ = 42(5)$~\unit{\micro\second} and $\tau_- = 54(4)$~\unit{\micro\second}, respectively. 
The two $\Delta m_s = \pm 1$ spin transitions  of the G centers are observed in ODMR spectra with a PL contrast of a few percents, constrained by a MS population detrapping mechanism induced by the above-bandgap excitation.
A level-anticrossing in the MS spin energy level structure of the G centers is evidenced at \SI{38.8\pm 0.4}{\unit{\milli\tesla}} by measuring the fluorescence intensity of the defects.  
The electron spin of G centers can be coherently manipulated, with a coherence time $T_2^\text{echo}=$~\SI{2.1\pm0.2}{\micro\second}, limited by the lifetime of short-lived MS spin state $|0\rangle$.

These results delineate clear directions for future work.
A central challenge will be to reduce or bypass the photo-detrapping channel, for instance by moving from above-bandgap to resonant optical excitation. 
This could enable the enhancement of the MS population, improving spin initialization and readout, and ultimately reaching higher-fidelity spin control.
An additional exploration path is to use the G center electron spin as ancillary qubit to control and detect nuclear spins, either at their vicinity of intrinsic to the defect atomic structure ($^{29}$Si or $^{13}$C; $I=1/2$, 4.7\% and 1.1\% at natural abundance, resp.)~\cite{akhtar_coherent_2012,lee_readout_2013}.
As electron spin flips are the main source of nuclear spin decoherence in materials purified with spin-less isotopes~\cite{taminiau_detection_2012}, the G center energy level structure could provide long-lived quantum memories based on individual nuclear spins, potentially approaching the hour-long coherence times measured for ionized donors in $^{28}$Si \cite{saeedi_room-temperature_2013}.
Finally, this work opens a new route towards optically-interfaced spin quantum memories and spin quantum registers \cite{song_entanglement_2026} within the mature silicon CMOS ecosystem~\cite{sandholzer_single-photon_2026}.

\section*{Acknowledgments}
This work is supported by  the Plan France 2030 through the project OQuLuS (No.~ANR-22-PETQ-0013), the French National Research Agency (ANR) through the project WOUAH (No.~ANR-24-CE47-4667), CEA through the PTC-MP “W-TeQ”  internal project, and the European Research Council (ERC) under the European Union’s  Horizon 2020 research and innovation programme (project SILEQS, Grant No.~101042075).  
Additional support is acknowledged from the Research Council of Norway through the DSQS project (project number 354831).
The authors thank Jean-Michel Hartmann for supplying the $^{28}$SiOI sample and Margriet van Riggelen for her contributions to the experimental developments.

	\bibliography{bib_G_ens}

	\end{document}